\documentclass[prl,twocolumn,amsmath,amssymb,superscriptaddress, nofootinbib,aps,10pt]{revtex4-1}

\usepackage{amssymb}
\usepackage{bbold}
\usepackage{xcolor}
\usepackage{makecell}
\usepackage{enumerate}
\usepackage[shortlabels]{enumitem}
\usepackage{graphicx}
\usepackage[caption=false]{subfig}

\usepackage{hyperref}
\usepackage[capitalise]{cleveref}

\newcommand{\vs}{\emph{vs}~}
\newcommand{\ie}{i.e.~}

\begin{document}

\title{Scaling description of creep flow in amorphous solids}

\author{Marko Popovi{\' c}}
\affiliation{Institute of physics, EPFL, Lausanne}
\affiliation{Max Planck Institute for Physics of Complex Systems, Nöthnitzer Strasse 38, 01187 Dresden, Germany}
\affiliation{Center for Systems Biology Dresden, Pfotenhauer Str. 108, 01307, Dresden, Germany}

\author{Tom W.~J.~de Geus}
\affiliation{Institute of physics, EPFL, Lausanne}

\author{Wencheng Ji}
\affiliation{Institute of physics, EPFL, Lausanne}

\author{Alberto Rosso}
\affiliation{LPTMS,CNRS,Univ.Paris-Sud,Universit{\' e} Paris-Saclay, 91405 Orsay, France}

\author{Matthieu Wyart}
\affiliation{Institute of physics, EPFL, Lausanne}

\begin{abstract}
Amorphous solids such as coffee foam, toothpaste or mayonnaise
display a transient creep flow when a stress $\Sigma$ is suddenly imposed.
The associated strain rate is commonly found to decay in time as $\dot{\gamma} \sim t^{-\nu}$,
followed either by arrest or by a sudden fluidisation.
Various empirical laws have been suggested for the creep exponent $\nu$ and
fluidisation time $\tau_f$ in experimental and numerical studies.
Here, we postulate that plastic flow is governed by the difference between $\Sigma$ and 
the transient yield stress $\Sigma_t(\gamma)$ that characterises the stability of configurations visited by the system at strain $\gamma$.
Assuming the analyticity of $\Sigma_t(\gamma)$ allows us to predict $\nu$ and asymptotic behaviours of $\tau_f$
in terms of properties of stationary flows.
We test successfully our predictions using elastoplastic models and published experimental results.
\end{abstract}

\maketitle

Amorphous materials including atomic glasses, colloidal suspensions,
dense emulsions or foams are important in industry and engineering \cite{Andreotti2013,Bonn2017}.
From a fundamental viewpoint, their properties are mesmerizing:
(i) Under quasi-static loading they
can display an avalanche-type response \cite{Maloney2006}
near their yield stress $\Sigma_c$. 
(ii) For $\Sigma>\Sigma_c$, they can present a singular flow curve, corresponding to the so-called Herschel-Bulkley's law \cite{Herschel1926} where the strain rate follows
$\dot{\gamma} \approx  c (\Sigma - \Sigma_c)^{\beta}$  with $c$ a material-specific constant and $\beta > 1$, see e.g. \cite{Lin2014b}. We restrict ourselves to  materials with such flow curves.
(iii) Depending on the system preparation the transient response to an applied strain
can be smooth, or discontinuous if a narrow shear band appears
 \cite{Ozawa2018,Popovic2018,Fielding2021}.
Here we focus on (iv) {\it creep flows}, another transient phenomenon observed when
a constant stress $\Sigma$ is imposed at time $t=0$ on an initial state at zero applied stress.
Transiently, a flow rate $\dot{\gamma}\sim t^{-\nu}$ is observed.
At low $\Sigma$, flow eventually arrests.
However, at sufficiently high $\Sigma$, $\dot\gamma(t)$ can be non-monotonic:
a sudden fluidisation may occur at some time $\tau_f$.
Commonly, the creep flow exponent $\nu$ is measured preceding the fluidisation
and reported in the range $0.34 - 1.2$
in experiments \cite{Bauer2006, Caton2008, Divoux2011, Siebenbuerger2012, Grenard2014, Leocmach2014} and
particle simulations \cite{Chaudhuri2013, Landrum2016, Cabriolu2019}.
By contrast,
the creep flow arrest is much less studied \cite{Siebenbuerger2012},
and $\tau_f$ is often reported using phenomenological fitting functions,
including:
(a) A power law $\tau_f \sim (\Sigma - \Sigma_0)^{-b}$
(with both $b$ and $\Sigma_0$ fitting parameters)
in experiments on carbopol microgel \cite{Divoux2011},
protein gels \cite{Leocmach2014}, and colloidal glasses \cite{Siebenbuerger2012};
and particle simulations \cite{Chaudhuri2013}.
(b) An exponential $\ln \tau_f \sim -\Sigma$
in experiments on carbon black gels \cite{Gibaud2010, Grenard2014} and silica gels \cite{Sprakel2011}.

From a computational viewpoint, studies of creep flow in
athermal elastoplastic models \cite{Nicolas2018}
report (a)
$\tau_f \sim (\Sigma - \Sigma_0)^{-b}$ with a preparation-dependent exponent
$b \simeq 1.7 - 2.2$ in a two-dimensional model \cite{Liu2018b} and
$b \simeq 1.3 - 2.2$ in a mean-field model \cite{Liu2018a}.
At finite temperature, both models are consistent with
(b) $\ln \tau_f \sim -\Sigma$ \cite{Merabia2016}.
The creep exponent $\nu$ was observed to be unity \cite{Bouttes2013}
or to be preparation dependent \cite{Merabia2016}.
Theoretical approaches supporting
particular fitting choices are mostly lacking.
A notable exception is the continuum model of shear banding \cite{Benzi2019}
that proposes $b= 9\beta/4$.

Here, we introduce a theoretical framework that predicts the exponent $\nu$, the asymptotic properties of $\tau_f$,
and their dependence on temperature. We focus on long time scales and assume that flow is then essentially plastic, thus neglecting the elastic contribution to the strain.
We expect this assumption to hold in the materials we consider here, coined ``simple
yield stress fluids" \cite{ovarlez2013rheopexy} such as foams, emulsions or repulsive colloidal glasses.  It does not hold in materials with a very slow linear visco-elastic response that can contribute to creep \cite{Aime2018, Aime2018b,Leocmach2014}. We also exclude loosely connected colloidal gels, which can display non-monotonic flow curves and sudden transition between distinct structures \cite{lindstrom2012structures,cho2021yield}. 
Our central hypotheses are that the plastic flow is governed by $\Sigma - \Sigma_t(\gamma)$,
where $\Sigma_t(\gamma)$ is a smooth function of plastic strain $\gamma$ that characterises the stability of configurations visited by the system at a strain $\gamma$.
These assumptions lead to a comprehensive description of creep flows in terms of
the Herschel-Bulkley exponent $\beta$,
as is summarised in \cref{tab:I} for athermal and \cref{tab:II} for thermal systems.
We confirm our predictions in two-dimensional and mean field elastoplastic models.
We find that our athermal predictions are also in good agreement with experiments on carbopol microgel and colloidal glasses,
while our thermal predictions are consistent with experiments on kaolin suspensions and ketchup.

\begin{figure}
    \subfloat{\label{fig:1:a}}
    \subfloat{\label{fig:1:b}}
    \subfloat{\label{fig:1:c}}
    \subfloat{\label{fig:1:d}}  
    \centering
    \includegraphics[width=\linewidth]{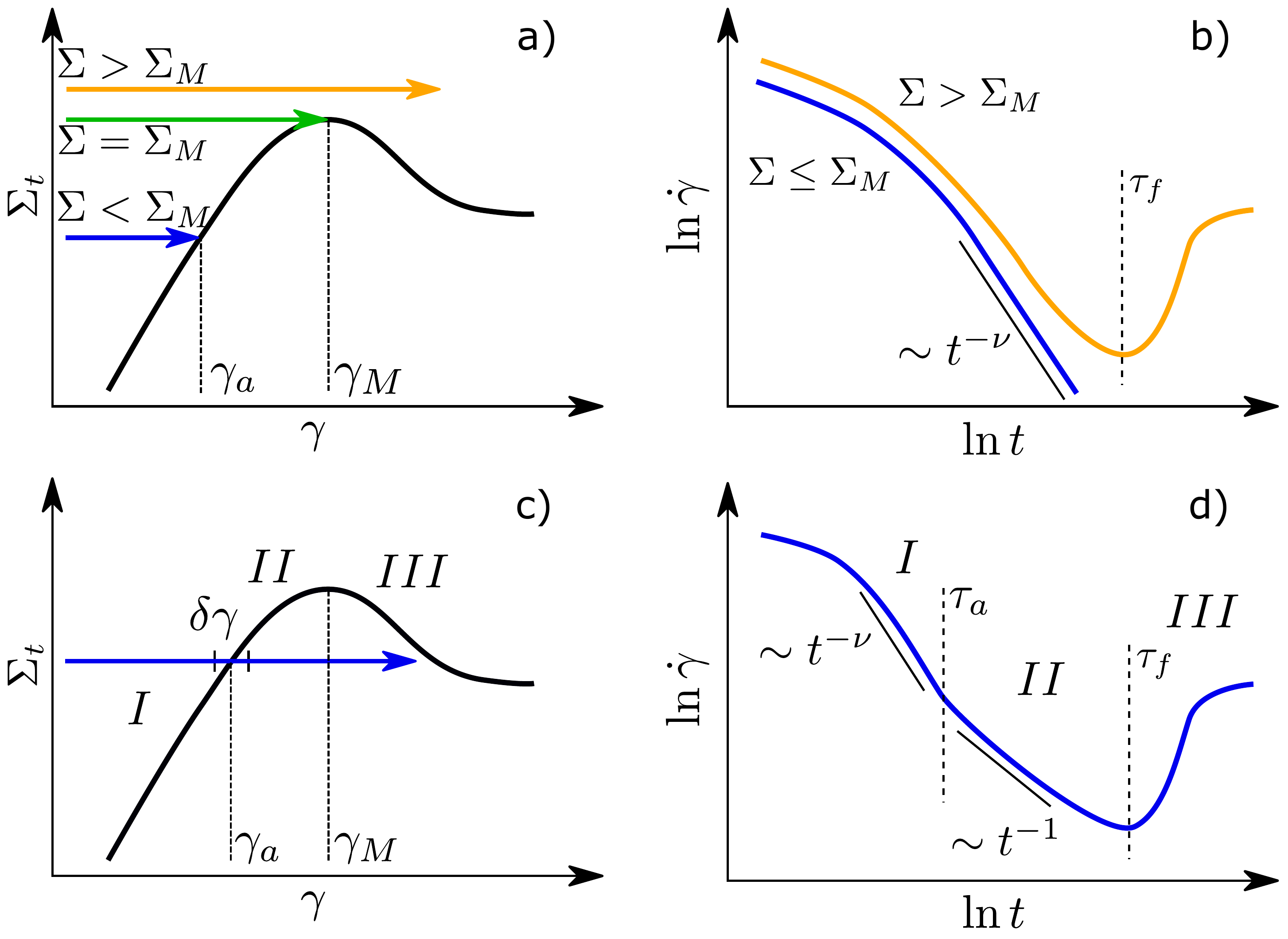}
    \caption{
        Left: Sketch of $\Sigma_t(\gamma;\Sigma,T)$
        for (a) $T = 0$ and (c) $T > 0$. Arrows indicate different applied stresses $\Sigma$ that lead to creep flow scenarios discussed in the text.
        Right: The corresponding sketch of the creep flow, respectively in (b) and (d).
        $\gamma_a(\Sigma)$ is defined by $\Sigma = \Sigma_t(\gamma_a,\Sigma,T)$ and $\gamma_M= \gamma_a(\Sigma_M)$.
    }
\end{figure}

{\it Theory:} The transient response of amorphous materials strongly depends on preparation.
For example, the quasistatic stress \vs plastic strain curve
can increase monotonically or overshoot \cite{Andreotti2013, Antonaglia2014, Ozawa2018} as the stability of the system preparation increases.
During quasistatic loading the system is at the stress which the material can withhold without flowing at plastic strain $\gamma$.
Here, we define the \textit{transient yield stress} $\Sigma_t(\gamma; \Sigma, T)$ that characterizes the stability of the material for non-quasistatic loading. At zero temperature $T$, its definition is:
\begin{align}
    \label{eq:1}
    \dot{\gamma} \equiv c \left( \Sigma - \Sigma_t(\gamma; \Sigma, T = 0) \right)^{\beta} \quad ,
\end{align}
To lighten notations, when possible we omit the dependence of $\Sigma_t$   on $\Sigma$ and $T$ and simply write it $\Sigma_t(\gamma)$. From Eq. (\ref{eq:1}), it follows that the flow  arrests at the finite strain $\gamma_a$ for which $\Sigma_t(\gamma_a) = \Sigma$, while in the steady state $\Sigma_t (\gamma \to \infty)= \Sigma_c$.  Note that $\Sigma_t(\gamma)$  so defined can be measured  by observing the creep flow dynamics and inverting Eq. (\ref{eq:1}), as performed below. 
Our central result is that simply assuming that $\Sigma_t(\gamma)$ is a smooth is sufficient to determine the creep flow exponent $\nu$ and the fluidisation time $\tau_f$, see \cref{fig:1:b}.
$\Sigma_t(\gamma)$ in general depends on the preparation of the system, similar to the quasistatic stress \vs strain curve. Here, we focus on the creep flow in systems where $\Sigma_t(\gamma)$ overshoots to a maximal value $\Sigma_M$ before reaching its steady state value $\Sigma_c$, as illustrated in \cref{fig:1:a}.  The case where $\Sigma_t(\gamma)$ does not overshoot, and instead grows monotonically can be treated with the same arguments. As shown in the Supplemental Material (SM), the strain rate monotonically decreases to the steady state value.

At low imposed stresses $\Sigma < \Sigma_M$, the flow arrests at a finite $\gamma_a$ (see \cref{fig:1:a}) where $\Sigma_t(\gamma_a) = \Sigma$.  By expanding $\Sigma_t \simeq \Sigma_t(\gamma_a) + \partial_{\gamma}\Sigma_t(\gamma_a) (\gamma - \gamma_a)$ and using Eq. (\ref{eq:1}), one obtains $\dot\gamma\sim (\gamma_a-\gamma)^\beta$ implying $\dot{\gamma} \sim t^{-\beta/(\beta - 1)}$.
Instead, for $\Sigma=\Sigma_M=\max_\gamma\Sigma_t(\gamma)=\Sigma_t(\gamma_M)$, a second order expansion implies that $\Sigma_t(\gamma)\approx\Sigma_M+ \partial^2_{\gamma}\Sigma_t(\gamma_{M}) (\gamma-\gamma_M)^2/2$.  Using again Eq. (\ref{eq:1}) one gets $\dot\gamma\sim (\gamma_M-\gamma)^{2\beta}$ and therefore  $\dot{\gamma}\sim t^{-2\beta/(2\beta - 1)}$. Finally, for $\Sigma > \Sigma_M$ the flow transiently slows down, reaching its minimum at $\gamma_M$. In the vicinity of $\gamma_M$, one has $\dot\gamma\sim [\Sigma-\Sigma_M+ \partial^2_{\gamma}\Sigma_t(\gamma_{M}) (\gamma_M-\gamma)^2/2]^{\beta}$.
The fluidisation time $\tau_f$ is the time at which $\gamma_M$ is reached. It is dominated by the time spent approaching $\gamma_M$ in an interval of strain of order $\Delta \gamma \sim (\Sigma-\Sigma_M)^{1/2} $, at a pace $\dot\gamma\sim (\Sigma-\Sigma_M)^\beta$, leading to a time
   $\tau_f \sim \Delta \gamma/\dot\gamma\sim \left( \Sigma - \Sigma_M \right)^{1/2 - \beta}$.
We summarise the athermal creep flow results in \cref{tab:I}.

\begin{table}
    \begin{tabular}{|c||c|}
        \hline
        ${\Sigma < \Sigma_M}$ & $\nu= \frac{\beta}{\beta-1}$ \\
        \hline
        ${\Sigma = \Sigma_M}$ & $\nu= \frac{2\beta}{2\beta - 1}$ \\
        \hline
        ${\Sigma > \Sigma_M}$ & ${\tau_f \sim (\Sigma - \Sigma_M)^{\frac{1}{2} - \beta}}$ \\
        \hline
    \end{tabular}
    \caption{
        Main results for athermal creep flow, illustrated in \cref{fig:1:a}.
        The corresponding creep flow scenarios are illustrated in \cref{fig:1:b},
        and corresponding numerical tests are shown in \cref{fig:2}.
    }
    \label{tab:I}
\end{table}

For a small finite temperature $T$ \footnote[1]{
    Corresponding to $T\ll T_g$, where $T_g$ is the glass transition temperature.
}, $\Sigma_t(\gamma; \Sigma, T)$ can now be defined from the finite temperature stationary flow curves. Our qualitative results are robust to details of the functional form chosen for these curves. Quantitatively, theoretical arguments and elastoplastic models
\cite{Chattoraj2010, Purrello2017, Ferrero2021, Popovic2021b}
support that the steady state flow follows a scaling relation:
$\dot{\gamma} =T^{\psi}f((\Sigma-\Sigma_c)/T^{1/\alpha})$.
Here, $\psi= \beta/\alpha$, where the parameter $\alpha$
describes the microscopic potential\footnote[2]{
    The exponent $\alpha$ characterises how the energy barrier $\Delta E$
    associated to a plastic event depends on the
    additional stress $\Delta \Sigma$ needed to trigger it, as
    $\Delta E\sim \Delta \Sigma^\alpha$.
For smooth interaction potentials between particles, plastic rearrangements correspond to saddle node bifurcations and $\alpha= 3/2$. For a potential with cusps $\alpha= 2$, as occurs for example in foams or in the vertex model of tissues \cite{Popovic2021}.}.
The scaling function $f$ must be such that $\dot{\gamma}$
converges to $\dot{\gamma} \sim (\Sigma - \Sigma_c)^{\beta}$
(the Herschel-Bulkley law) in the limit $T\to 0$,
\ie $f(x)\sim x^\beta$ for $x\rightarrow \infty$.
For negative arguments, $f$ describes thermal activation so that
$f(x)\sim \exp(-C_0 x^\alpha)$ for $x\rightarrow - \infty$, where $C_0 > 0$.

We thus define the transient yield stress at finite $T$ as:
\begin{align}
    \label{eq:2}
    \dot{\gamma} \equiv T^{\psi} f \left( \frac{\Sigma - \Sigma_t(\gamma; \Sigma, T)}{T^{1/\alpha}} \right) \quad .
\end{align}
Here we discuss systems where $\Sigma_t(\gamma)$ overshoots, as illustrated in \cref{fig:1:c,fig:1:d}, see SM for the monotonic case, which includes Ref. \cite{Lin2015}. Initially at small strains thermal fluctuations are negligible and the creep flow exponent follows the athermal prediction $\dot\gamma\sim t^{-\beta/(\beta-1)}$. 
This regime is valid until a plastic strain $\gamma_a$ for which $\Sigma \simeq \Sigma_t(\gamma_a)$, where Eq. (\ref{eq:2}) implies that the flow rate  follows $\dot{\gamma} \sim T^{\psi}$. Comparing these two expressions,
the crossover time  where thermal activation starts to play a role follows $\tau_a \sim T^{(1-\beta)/\alpha}$. This cross-over occurs on a strain increment $\delta\gamma$ (see \cref{fig:1:c}), which corresponds to the argument of $f$ in Eq. (\ref{eq:2}) becoming negative and ${\cal O}(1)$. Expanding this argument using $\Sigma -\Sigma_t(\gamma)\sim \gamma_a-\gamma$ leads to $\delta\gamma \sim T^{1/\alpha}$. 
Beyond the crossover $\gamma - \gamma_a \gg \delta \gamma$, flow is dominated by thermal activation. This corresponds to the exponential behaviour of $f(x)$ for large negative arguments. It is then straightforward (see SM) to obtain from Eq. (\ref{eq:2}) and the linearization $\Sigma -\Sigma_t(\gamma)\sim \gamma_a-\gamma$  that the strain  grows logarithmically in time, implying that $\dot{\gamma} \sim t^{-1}$ at  long times.
Finally, for $\gamma > \gamma_M$ the flow rate rises and fluidisation occurs.
In contrast to athermal systems,  fluidisation also occurs for $\Sigma < \Sigma_M$.
We can estimate the fluidisation time in the limit of small temperatures,
as the time spent in the vicinity of $\gamma_M$.
For $\Sigma < \Sigma_M$, expanding $\Sigma_t(\gamma)$ around $\gamma_M$ in Eq. \ref{eq:2} and using the scaling function form we derived previously \cite{Popovic2021b}, we find $\tau_f \sim (T/(\Sigma_M - \Sigma)^{\alpha - 1})^{1/2 - \beta} \exp[(C_0 (\Sigma_M - \Sigma)^{\alpha}/T)]$.
For $\Sigma > \Sigma_M$ the flow is predominantly athermal,
except for $(\Sigma- \Sigma_M)^\alpha\leq T$ where  $\dot\gamma\sim T^\psi$ for strains near $\gamma_M$ on an interval that scales as $\Delta\gamma=\gamma_M-\gamma\sim T^{1/(2\alpha)}$,
leading to a fluidisation time  $\tau_f \sim \Delta\gamma/\dot\gamma\sim T^{(1/2 - \beta)/\alpha}$.

\begin{figure}[ht!]
    \subfloat{\label{fig:2:a}}
    \subfloat{\label{fig:2:b}}
    \subfloat{\label{fig:2:c}}
    \subfloat{\label{fig:2:d}}
    \centering
    \includegraphics[width=\linewidth]{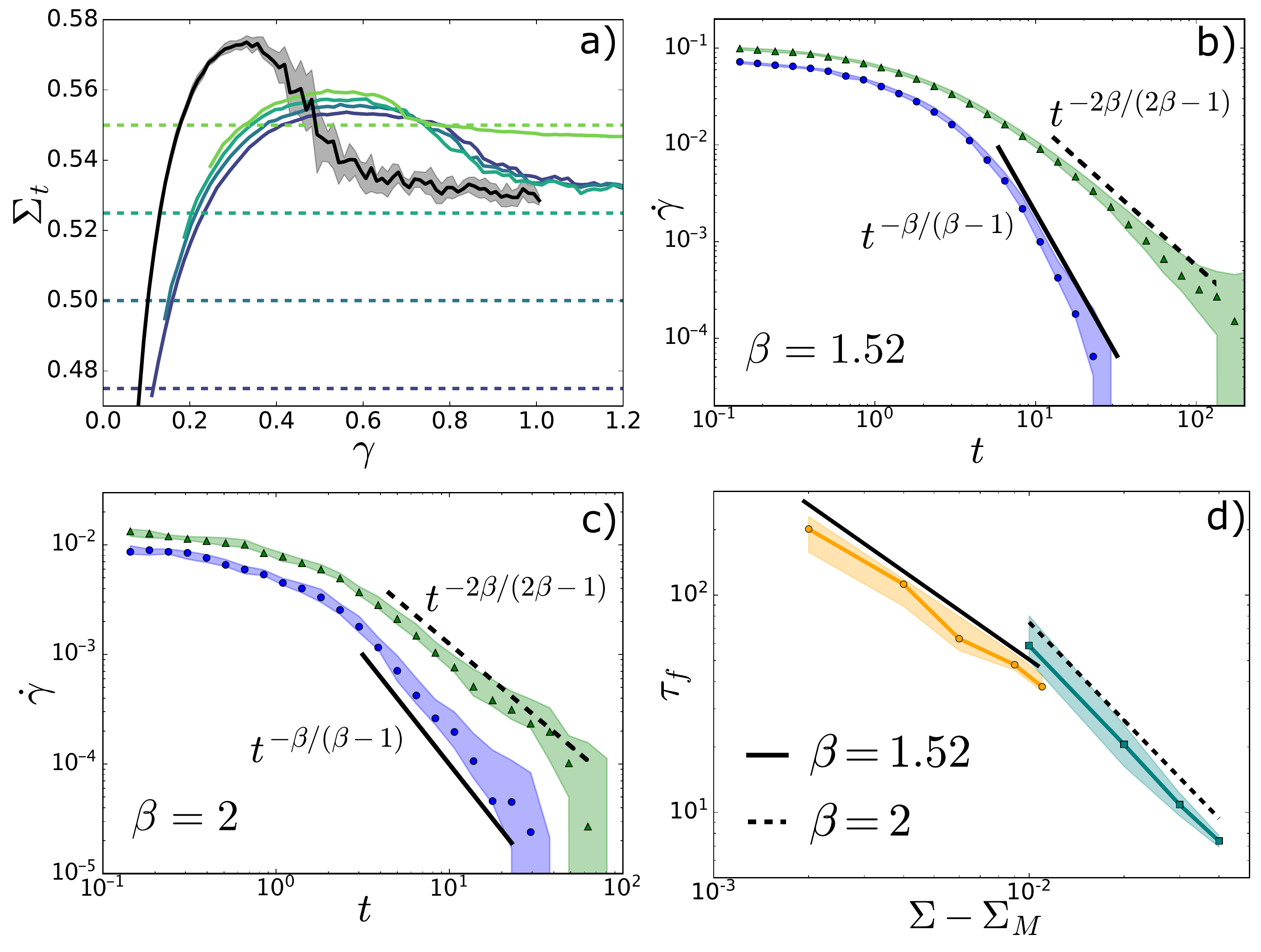}  
    \caption{Creep flow in athermal elasto-plastic models:
        (a) Transient yield stress curves $\Sigma_t(\gamma)$ 
        at stresses indicated by the dashed lines, at $T=0.002$.
        For reference,
        the quasistatic stress \vs plastic strain curve is shown in black.
        (b, c) Median values of creep flow 
        at two imposed stresses:
        blue circles ($\Sigma < \Sigma_M$),
        green triangles ($\Sigma = \Sigma_M$).
        Black lines indicate the corresponding predicted power laws, see \cref{tab:I}. 
        (d) Fluidisation times in $2d$ (yellow circles) and mean field
        (cyan squares).
        In all plots shaded regions correspond to $25\text{th}$-$75\text{th}$ percentile range.
    }
    \label{fig:2}
\end{figure}

\begin{table}
\resizebox{.48\textwidth}{!}{%
    \begin{tabular}{|c||c|}
        \hline
        \makecell{athermal to thermal\\ transition width} & $\delta \gamma \sim T^{1/\alpha}$\\
        \hline
        \makecell{athermal to thermal\\ transition time}  & ${\tau_{a} \sim T^{(\beta - 1)/\alpha}}$ \\
        \hline
        thermal creep flow & $\nu = 1$\\
        \hline
        fluidisation time &
        $\tau_f \sim (\frac{T}{(\Sigma_M- \Sigma)^{\alpha- 1}})^{\frac{1}{2} - \beta} e^{ c_T \frac{(\Sigma_M - \Sigma)^\alpha}{T}}$ \\
        \hline
    \end{tabular}}
    \caption{
        Main results for thermal creep flow. 
        The corresponding numerical tests are shown in \cref{fig:3}.
    }
    \label{tab:II}
\end{table}

{\it Numerical simulations:}
To test the proposed creep exponents we simulate creep flow using a two-dimensional
elastoplastic \cite{Popovic2021b} (see SM, which includes Ref. \cite{Picard2005}), whereby we benefit for previously
measured exponent $\beta = 1.52$ \cite{Lin2014b} and scaling function $f$ \cite{Popovic2021b}.

To estimate the athermal transient yield stress function $\Sigma_t(\gamma;\Sigma, T=0)$, we measure $\dot\gamma(t)$ at a tiny temperature $T= 0.002$ and then numerically invert \cref{eq:2} using the
previously measured $f$ \cite{Popovic2021b}, as shown in \cref{fig:2:a}.
We use a tiny but finite temperature to probe $\Sigma_t$ beyond the strain $\gamma_{a}$ at which athermal creep would arrest.
We find that $\Sigma_{t}$ changes with $\Sigma$, but this dependence is weak.
More importantly, our observations are consistent with our smoothness assumption.
For comparison, we show the quasistatic stress \vs plastic strain curve in the same system, which is clearly different from $\Sigma_t(\gamma)$.

We simulate the athermal creep flow at stresses $\Sigma \leq \Sigma_M$, see \cref{fig:2:b}. The measured creep flow dynamics is consistent with predictions summarised in \cref{tab:I}.
To further test our predictions, we use a mean-field version of elastoplastic model \cite{Popovic2021}, which
corresponds to a version of H\'ebraud-Lequeux model \cite{Hebraud1998}
where $\beta= 2$.
We again find that creep flow dynamics is consistent with our predictions, see \cref{fig:2:c}.

Finally, for imposed stresses $\Sigma > \Sigma_M$ we measure the fluidisation times $\tau_f$ as a function of the imposed stress $\Sigma$ in both models, as shown in \cref{fig:2:d}.
Although the range of data is less than a decade, the changes in the asymptotic behaviour of $\tau_f$ are consistent with our predictions, for both values of $\beta$.

\begin{figure}[ht!]
    \subfloat{\label{fig:3:a}}
    \subfloat{\label{fig:3:b}}
    \subfloat{\label{fig:3:c}}
    \subfloat{\label{fig:3:d}}
    \centering
    \includegraphics[width=\linewidth]{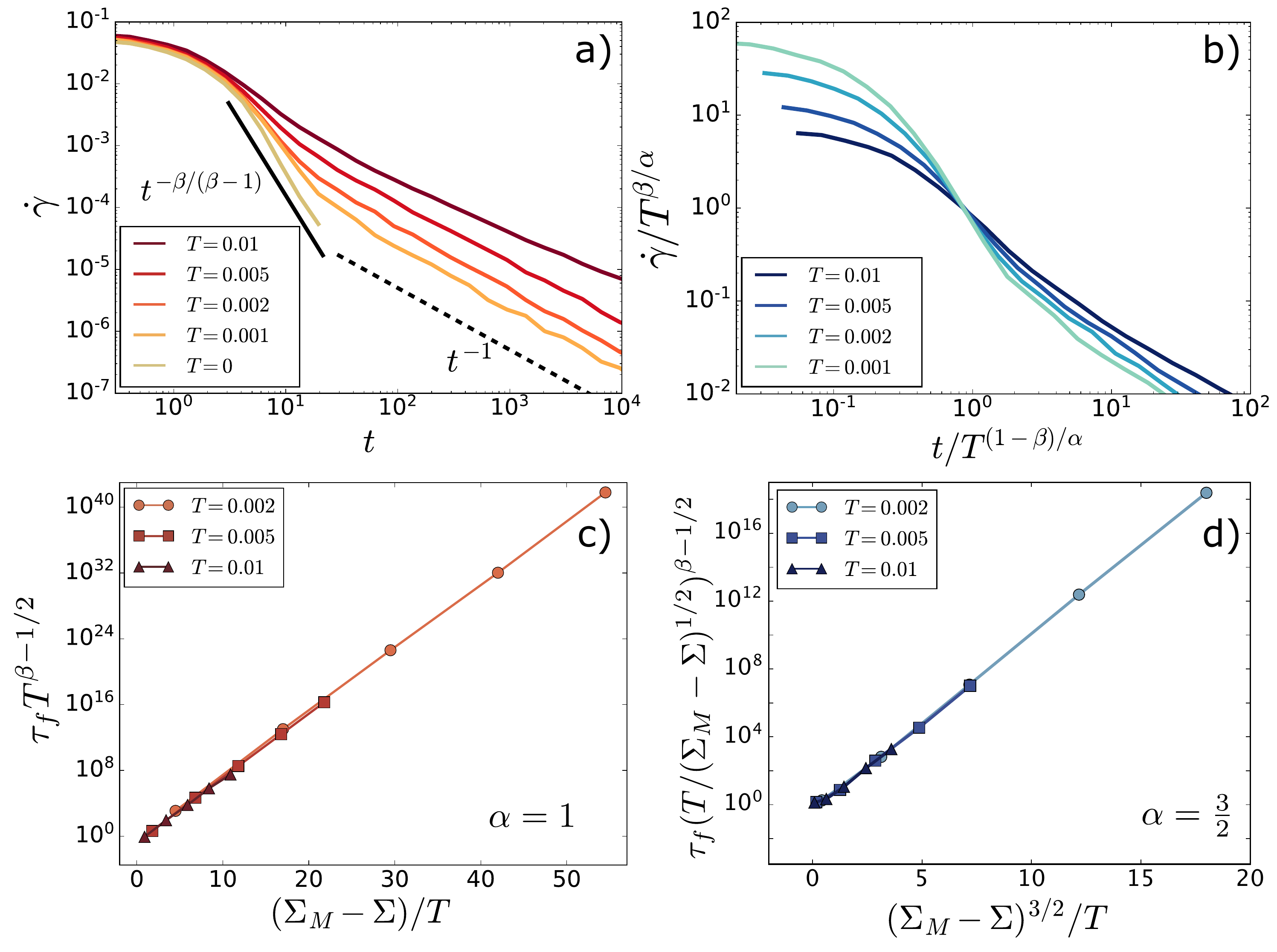}
    \caption{
        Creep flow in thermal $2d$ elastoplastic model:
        (a) Athermal and thermal creep regimes follow the predicted flow rate exponents.
        (b) Rescaling flow rate and time collapses the crossing point of all curves,
        confirming the existence of a crossover time scale  $\tau_a \sim T^{(1 - \beta)/\alpha}$.
        (c, d) Fluidisation times $\tau_f$ measured with $\alpha=1$ (c)
        and $\alpha= 3/2$ (d) at different temperatures are consistent with our prediction.
    }
    \label{fig:3}
\end{figure}

We next turn to thermal systems. We first study the transition from the athermal to the thermal creep regime,
sketched in \cref{fig:1:c,fig:1:d}.
In \cref{fig:3:a} we show creep curves for $\alpha= 3/2$ at $\Sigma= 0.45$ in a system with an overshoot in $\Sigma_t(\gamma)$.
As the temperature is decreased towards $T= 0$, the transition between the athermal regime
($\dot{\gamma} \sim t^{-\beta/(\beta - 1)}$) and thermal creep ($\dot{\gamma} \sim t^{-1}$)
is indeed observed, and occurs at later times following $T^{(1 - \beta)/\alpha}$,
as confirmed in \cref{fig:3:b}.

Finally, we measure fluidisation times
of thermal creep flow at different temperatures and
imposed stresses both for $\alpha= 1$ (\cref{fig:3:c}) and $\alpha= 3/2$ (\cref{fig:3:d}).
Following \cite{Caton2008}, we define the fluidisation time as the time corresponding to the minimum of the flow rate.
We find an excellent collapse of the data, confirming our prediction
$\tau_f \sim (T/(\Sigma_M - \Sigma)^{\alpha - 1})^{1/2 - \beta} \exp[(C_0 (\Sigma_M - \Sigma)^{\alpha}/T)]$.

Note that our theory predicts asymptotic fluidisation and creep exponents in the limit of vanishing flow.
Therefore, the effective values extracted from the whole range of measured fluidisation times will in general differ from our measurements.
This could account for the differences to the preparation dependent effective exponents
reported in the extensive numerical simulations of athermal creep in elastoplastic models \cite{Liu2018b}.

\begin{figure}[ht!]
    \subfloat{\label{fig:4:a}}
    \subfloat{\label{fig:4:b}}
    \centering
    \includegraphics[width=\linewidth]{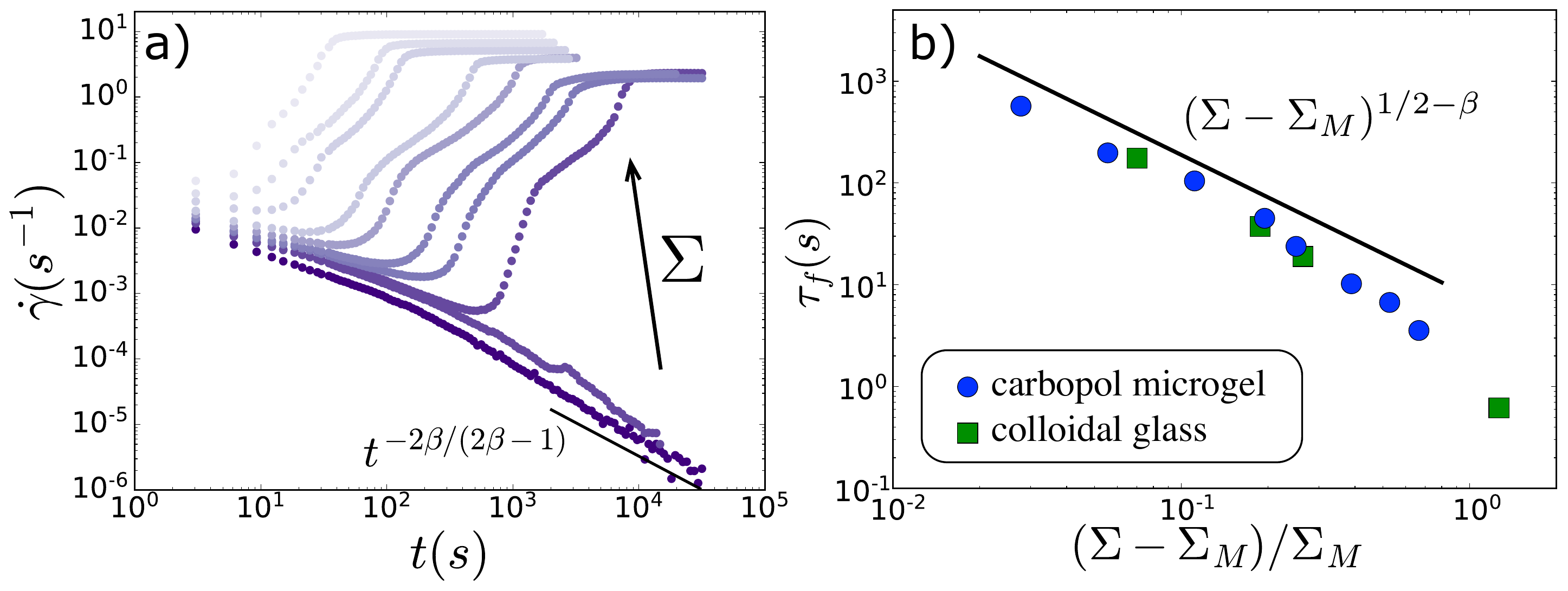}
    \caption{(a) Creep flow of carbopol microgel \cite{Divoux2011}
        at:
        $\Sigma[\text{Pa}]=35,36,37,38,40,43,45,50,55,60$ (from bottom to top).
        The arresting curves are consistent
        with our prediction (black line).
        (b) Fluidisation times (see main text for measurement) of
        carbopol microgel \cite{Divoux2011} (blue circles) and
        colloidal glass \cite{Siebenbuerger2012} (green squares)
        together with our prediction.
    }
    \label{fig:4}
\end{figure}

{\it Experimental tests:} We compare our results the experimental data from
carbopol microgel creep experiments \cite{Divoux2011}, reproduced in \cref{fig:4:a}. 
At imposed stress values just below the fluidisation stress,
the creep exponent is consistent with our prediction $\nu = 2 \beta / (2 \beta - 1)$,
where we use $1 / \beta = 0.53$ measured by \cite{Divoux2011}.
We then extract the fluidisation times from the minima of the flow curves both in this experiment
and in the colloidal glass experiment of \cite{Siebenbuerger2012}. 
As shown in \cref{fig:4:b},
it is consistent with our athermal prediction\footnote[3]{
    The steady state flow is reported to follow the Herschel-Bulkely law
    and therefore we expect the athermal regime to be relevant.
}
$\tau_f \sim (\Sigma - \Sigma_M)^{1/2-\beta}$, as indicated by the black line, where the value of
$\Sigma_M$ is estimated as the highest reported stress value
for which no fluidisation is observed, 
and we use $\beta= 1.89$ from \cite{Divoux2011}.

Note that another definition of fluidization time $\tau_f^*$, corresponding to the inflection point of the creep curve, was used in \cite{Divoux2011, Gibaud2010, Grenard2014}. $\tau_f^*$  is associated with the emergence of shear banding \cite{Divoux2011,Benzi2019}. Our theory for fluidization, which assumes a homogeneous flow and does not capture shear banding, may thus apply as long $\tau_f\leq \tau_f^*$. This inequality is fulfilled in the cited examples, and also in theoretical treatment supporting that the flow remains homogeneous before $\tau_f$ \cite{Moorcroft2013}.

Concerning thermally activated creep flow, we predict an exponential
dependence of $\tau_f$ on $\Sigma$, which was indeed reported in carbon black gels
\cite{Gibaud2010,Grenard2014},
and in numerical simulations of thermally activated flow in elastoplastic models \cite{Merabia2016}.
Likewise, our prediction for the thermal creep flow regime $\dot{\gamma} \sim t^{-1}$
is found in numerical simulations of thermally activated flow \cite{Bouttes2013}. This behavior is also found in kaolin suspensions \cite{Uhlherr2005} and ketchup \cite{Caton2008}. However, the validity of our approach to these materials is less clear, as their flow curves need not follow a Herschel-Bulkley law as we assume. They can be instead thixotropic materials with non-monotonic flow curves \cite{ovarlez2009phenomenology}, known to shear band in  stationary flows.

{\it Discussion:}
We have provided a theoretical framework in which creep flows are controlled
by the stress $\Sigma_t$ at which configurations visited at time $t$ would stop flowing.
Our treatment is similar
in spirit to the Landau theory of a phase transition: assuming the analyticity of $\Sigma_t$
enables one to express the asymptotic behaviours of creep flows in terms of
the better understood stationary flows.
Our analysis predicts a rich set of regimes, which is consistent with observations
in elastoplastic models and in experiments.

Usual mean-field approaches, both for the yielding transition in
amorphous solids \cite{Hebraud1998,Lin2016a} and for the depinning transition \cite{Fisher1998},
consider the dynamics of the distribution $P(x)$, where $x$ is a local variable
indicating how much additional shear stress is required to have a plastic event.
In such models, rate of plastic activity following some initial condition was computed at
$\Sigma= 0$ and $T=0$ \cite{Sollich1997,Parley2020}.
These results are consistent with our prediction for $\nu$,
supporting that our assumption of analyticity is equivalent to mean-field approaches
as is the case in Landau theory.

Our assumption should thus
break down when spatial correlations are large, which occurs
in particular if avalanches are compact objects.
It is the case for short-range depinning phenomena if the spatial dimension satisfies $d<4$,
in that case an alternative real space scaling approach summarised in
the SM is needed, which includes Refs. \cite{Kolton2006, Ferrero2013}.
By contrast, we expect our analysis to hold if $d\geq 4$,
or in amorphous solids since in that case avalanches are not compact: the density of plastic events within them vanishes as the avalanche linear extension grows \cite{Lin2014b,Nicolas2018}.

\textbf{Acknowledgments}
T.G.~acknowledges support from The Netherlands Organisation for Scientific Research (NWO)
by a NWO Rubicon Grant 680-50-1520 and from the Swiss National Science Foundation (SNSF)
by the SNSF Ambizione Grant No.~PZ00P2\_185843.
The project was supported by the Simons Foundation Grant (No. 454953 Matthieu Wyart)
and from the SNSF under Grant No. 200021-165509.

\bibliographystyle{apsrev4-1}
\bibliography{mpopovicBib}

\end{document}


\title{Supplemental material for Scaling description of creep flow in amorphous solids}
\author{Marko Popovi{\' c}}
\address{Institute of physics, EPFL, Lausanne}
\address{Max Planck Institute for Physics of Complex Systems, Nöthnitzer Strasse 38, 01187 Dresden, Germany}
\affiliation{Center for Systems Biology Dresden, Pfotenhauer Str. 108, 01307, Dresden, Germany}
\author{Tom W. J. de Geus}
\address{Institute of physics, EPFL, Lausanne}
\author{Wencheng Ji}
\address{Institute of physics, EPFL, Lausanne}
\author{Alberto Rosso}
\address{LPTMS,CNRS,Univ.Paris-Sud,Universit{\' e} Paris-Saclay, 91405 Orsay, France}
\author{Matthieu Wyart}
\address{Institute of physics, EPFL, Lausanne}

\maketitle

\subsection{Coarsening dynamics}
The arguments of this paper, based on the assumption that a function is analytic, are in the spirit of  the Landau theory of phase transitions. They  are thus similar to a mean field approximation.
They  provide a consistent description of creep experiments because plastic avalanches are
collections of events that are very sparse in space.
This is not always the case.
For example, the avalanches at the depinning transition destabilize
compact portions of the elastic interface.
Thus, in these systems, the scaling  results based on the assumption of some function is analytic  cannot work, and should
be replaced by a description in terms of coarsening of domains where the interface is rough.
The latter approach correctly describes the dynamics of a system after a rapid quench
from a homogeneous phase to a critical point, or to a region of two phases
(e.g.\ a ferromagnetic region).
For a concrete illustration we consider an interface, initially flat,
suddenly pulled at the critical force of the depinning transition.
This protocol is analogous to imposing a stress $\Sigma_c =\Sigma_{\max}$
on an amorphous material at zero temperature.
However, for a $d-$dimensional interface, after a microscopic time scale,
a dynamical-scaling regime emerges in which the interface is rough
(with positive roughness exponent $\zeta$ for $d<4$) up to a coarsening scale, $\ell(t)$.
The coarsening length  grows in time  as $\ell \sim t^{1/z}$,
where $z>\zeta$ is the  dynamic exponent.
As a consequence, the interface's center of mass grows sublinearly with time as
$\sim \ell^{\zeta}  \sim t^{\zeta/z}$ and the interface's velocity slowly decays with time as
\begin{equation}
v \sim
t^{-( z-\zeta)/z}   \; \text{for} \; \zeta>0
\end{equation}
This behaviour has been well verified in \cite{Kolton2006, Ferrero2013}
and cannot be predicted using our approach.
The case where both approaches agree is
the  mean-field depinning  (for $d\ge 4$).
There  we still expect, at the depinning critical force,
an unbounded growth of the interface center of mass,
but logarithmic instead of power law, as $\zeta=0$.
In this case the velocity  decays as $1/t$ as also predicted by our arguments.

\subsection{Transiently inhomogeneous flow}
Our analysis should break down when flow is inhomogeneous and transient shear banding occurs,
since these phenomena are absent from stationary flows.
However, shear banding often occurs after the maximum in the stress \emph{vs} strain curve
(as appears to be the case in the carbopol microgel experiment described above),
in which case our predictions for $\tau_f$ should still hold.
Nevertheless, for extremely stable glasses, narrow shear bands can occur before that maximum is reached,
see e.g. \cite{Ozawa2018}.
In that case, the fluidisation time is likely controlled by the nucleation of a narrow shear band,
whose mechanism is debated \cite{ Ozawa2018,Popovic2018, Fielding2021}.

\subsection{Thermal elastoplastic model}\label{app:model}

We employ a two-dimensional elastoplastic model.
In this model the material is divided into $N$ mesoscopic blocks,
characterized by the local stress component $\sigma_i$ along the external loading direction,
the shear elastic modulus $\kappa$, and a local yield stress $\sigma_{Y, i}$.
When $|\sigma_i| > \sigma_{Y, i}$ block $i$ fails at a rate $1/\tau$.
As long as $|\sigma_i| < \sigma_{Y, i}$ block $i$ is stable in athermal systems,
while for $T > 0$ it fails at a rate
$\exp{\left((\sigma_{Y, i} - |\sigma_i|)^{\alpha})/T\right)}/\tau$.
After the failure stress is reduced by $\delta\sigma= \sigma_i + \mathcal{N}(0, 0.01)$,
i.e.\ the current stress with a small amount of noise added
to prevent possible periodic behaviour,
where $\mathcal{N}(\mu, s^2)$ is a normal distribution with mean $\mu$ and variance $s^2$.
The corresponding plastic strain $\delta\gamma_i= \delta\sigma_i/\kappa$
is accumulated in the block, see \cite{Popovic2021b} for further details.
Finally, each block failure redistributes stress in the system according to a propagator
$G(\vec{r})$ corresponding to a force dipole in elastic medium \cite{Picard2005, Nicolas2018}.
In this work we use an initial stress distribution in blocks that is distributed according to
a normal distribution $\mathcal{N}(0, 0.16)$.
The block yield stresses after a failure are drawn from a normal distribution
$\mathcal{N}(\mu, 0.01)$, with $\mu= 1$.
In order to generate a system where $\Sigma_t$ overshoots we choose $\mu_0= 1.1$,
and for a system that does not overshoot we choose $\mu_0= 1$.

In Fig.~2a (main text) we use the system with $\mu_0=1.1$ to compare the quasistatic loading curve with
the measurement of transient yield stress obtained by measuring transient
strain rate and inverting Eq.~2 (main text).

In mean-field simulations we always use a normal distribution of block yield stresses $\mathcal{N}(1, 0.01)$ and we distribute initial block stresses according to $\mathcal{N}(0, 0.16)$ and $\mathcal{N}(0, 0.01)$ to generate systems  without and with an overshoot in $\Sigma_t$, respectively.

\subsection{Creep flow for a monotonic transient yield stress}
Here we discuss the creep flow of a system in which the transient yield stress $\Sigma_t(\gamma)$ does not overshoot and monotonically rises to the steady state value $\Sigma_c$.

\subsubsection{Athermal systems: $T= 0$}

At low stresses $\Sigma < \Sigma_c$ the creep flow is the same as in the overshoot case presented in the main text: $\dot{\gamma} \sim t^{-\beta/(\beta - 1)}$. However, as the imposed stress approaches $\Sigma_c$ the linearisation at arresting strain $\gamma_a$ becomes insufficient and to discuss the creep flow for $\Sigma = \Sigma_c$ we assume an asymptotic form of $\Sigma_t(\gamma) \simeq \Sigma_c \left(1 - e^{-c_1\gamma}\right)$. This form is motivated by reported exponential asymptote of the quasistatic stress \vs strain curve in experiments \cite{Andreotti2013} and in elastoplastic models \cite{Lin2015}.
Such exponential asymptote can be understood as a consequence of plastic events erasing the
initial condition of the stress distribution in the material.
Now, it is straightforward to find  $\dot{\gamma}\sim t^{-1}$, independent of $\beta$\footnote{
            In case of a power-law asymptote
            $\Sigma_t(\gamma) \simeq \Sigma_c     \left( 1 - c \gamma^{-\eta} \right)$
            the creep flow would follow $\dot{\gamma} \sim t^{-\eta\beta/(1 + \eta\beta)}$.}.
Finally, for $\Sigma > \Sigma_c$   $\dot\gamma$ monotonically decreases toward its finite stationary value.

We have tested the predicted creep flow exponents for systems without overshoot in $\Sigma_t(\gamma)$ in $2d$ and mean-field simulations, shown in Fig. S1 b) and c), respectively. We find that, as for systems with an overshoot in $\Sigma_t(\gamma)$, our prediction of the creep flow exponent is consistent with numerically measured creep flow curves. 

\begin{figure}[ht!]
    \centering
    \includegraphics[width=.95\linewidth]{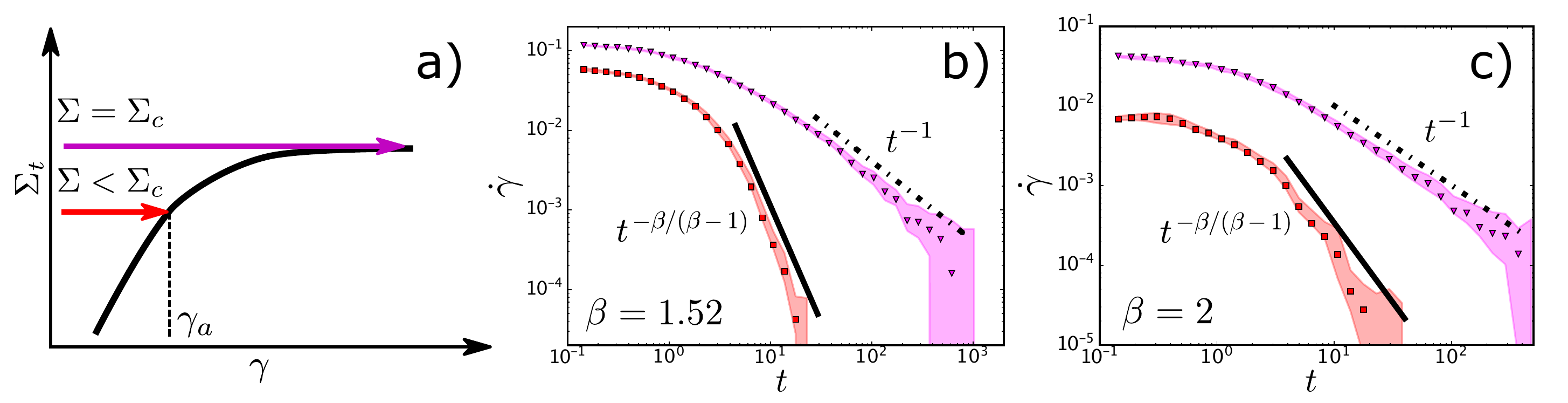}
    \caption{
    (a) Sketch of $\Sigma_t(\gamma; \Sigma, T=0)$ without an overshoot. Arrows indicate different creep flow scenarios discussed in the text.
    (b,c) Creep flow at two imposed stresses: $\Sigma < \Sigma_M$ (red squares), $\Sigma= \Sigma_M$ (magenta triangles) in $2d$ (b) and mean field (c). Black lines indicate the corresponding predictions discussed in the text.
    In both plots shaded regions indicate $25$-th and $75$-th percentile region while the markers are the median values of the creep flow.
    }
    \label{fig:noOvershootCreep}
 \end{figure}

\subsubsection{Thermal systems: $T > 0$}
In thermal system at imposed stress $\Sigma < \Sigma_c$ both initial athermal and subsequent thermal creep regimes presented in the main text for systems with an overshoot in $\Sigma_t(\gamma)$ exist also in the case without an overshoot in $\Sigma_t(\gamma)$. However, fluidisation does not occur and the flow monotonically decreases toward the steady state value.

\subsection{Thermal creep regime}
We estimate scaling of the crossover time between athermal and thermal regimes of thermal creep flow as the time spent in the athermal regime $\tau_{a, 1}$ plus time spent in the crossover $\tau_{a, 2}$.
The time in the athermal regime can be estimated as
$\tau_{a,1} \sim \int_0^{\gamma_a - \delta\gamma/2} (\gamma_a-\gamma)^{-\beta}d\gamma \sim \delta\gamma^{1-\beta} \sim T^{(1-\beta)/\alpha}$.
The time spent transversing the remaining crossover interval $\delta\gamma/2$
approaching the transition scales as
$\tau_{a,2} \sim \delta\gamma/T^{\psi} \sim T^{{(1-\beta)}/\alpha}$.
Therefore, $\tau_a= \tau_{a,1}+ \tau_{a,2} \sim T^{(1-\beta)/\alpha}$.

In the thermal regime the flow scaling function is of the form $\dot{\gamma} \sim T^{\psi}\exp[-C_0(\Sigma_t(\gamma; \Sigma, T) - \Sigma)^{\alpha}/T]$. To calculate the creep flow dynamics in the transient thermal regime after entering the thermal regime\footnote{Extending this result to larger strain occurring at extremely long time leads to additional log corrections.} we linearise $\Sigma_t\simeq \Sigma_t(\gamma_0) + c_2(\gamma - \gamma_0)$ 
around a reference strain $\gamma_0$ that satisfies $\gamma_0-\gamma_a \gg \delta\gamma$. 
Furthermore, for $\alpha \neq 1$ we linearise the expression in the exponent of the thermal flow rate
$(\Sigma_t(\gamma) - \Sigma)^{\alpha} \simeq (\Sigma_t(\gamma_0)-\Sigma)^\alpha + \alpha c_2 (\gamma - \gamma_0)(\Sigma_t(\gamma_0) - \Sigma)^{\alpha - 1}$.
Inserting these linearisations in the expression for the thermal flow we find
$\dot{\gamma} \sim T^{\psi} \exp{\left[-\tilde{C} (\gamma - \gamma_0)/T \right]}$,
where $\tilde{C}$ contains $C_0$ and prefactors from the linearisations.
This yields $\dot{\gamma} \sim (t + \tau_0)^{-1}$, where $\tau_0$ is an integration constant,
and therefore at long times $\dot{\gamma} \sim t^{-1}$.
In order for this scaling to be observable, the temperature has to be low enough
so that the linearisations in strain are valid over sufficiently long times.

\subsection{Athermal fluidisation time measurement}

We extract the fluidisation time $\tau_f$ from simulations with our elastoplastic models as
the median value of the plastic flow rate minima among $N= 8$ creep flow realisations that fluidise at a given stress $\Sigma$.
In Fig. S2 we show the creep flow curves that fluidise at different stresses in
$2d$ and mean field elastoplastic simulations.

\begin{figure}[ht!]
    \centering
    \includegraphics[width=.8\linewidth]{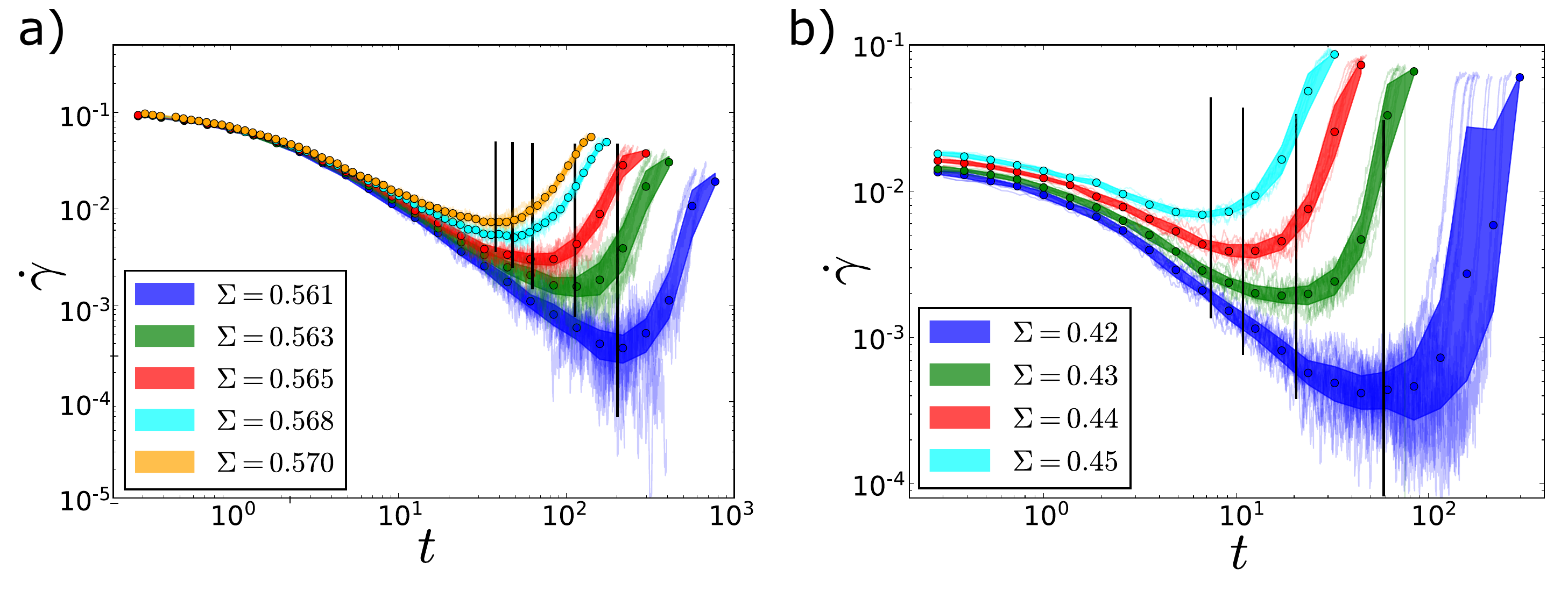}
    \caption{
        Creep flow curves from which the fluidisation times are extracted in
        Fig.~2 (main text) for $2d$ (a) and mean field (b) elastoplastic models.
        Shaded regions indicate $25$-th and $75$-th percentile region and circles
        the corresponding median values of the creep flow. Vertical black lines show
        the measured median values of $\tau_f$.
    }
    \label{fig:athermal}
  \end{figure}

\bibliographystyle{apsrev4-1}
\bibliography{mpopovicBib}